\documentclass[aps,showpacs,nofootinbib,amssymb,preprint]{revtex4}
\usepackage{graphicx,epsfig}
\def\bfnabla{\mbox{\boldmath $\nabla$}}

\def\bfsigma{\mbox{\boldmath $\sigma$}}

\newcommand{\nn}{\nonumber}
\newcommand{\be}{\begin{equation}}
\newcommand{\ee}{\end{equation}}
\newcommand{\bea}{\begin{eqnarray}}
\newcommand{\eea}{\end{eqnarray}}
\def\al{\alpha}

\def\siml{{\ \lower-1.2pt\vbox{\hbox{\rlap{$<$}\lower6pt\vbox{\hbox{$\sim$}}}}\ }} 

\newcommand{\MS}{\overline{\rm MS}}
\def\lQ{\Lambda_{\rm QCD}}

\begin{document}

\title{\boldmath 
Next-to-leading ultrasoft running of the heavy quarkonium potentials and spectrum: 
Spin-independent case
\unboldmath}
\author{Antonio Pineda}
\affiliation{Grup de F\'\i sica Te\`orica, Universitat
Aut\`onoma de Barcelona, E-08193 Bellaterra, Barcelona, Spain}

\date{\today}

\begin{abstract}
\noindent
We compute the next-to-leading logarithmic (NLL)  ultrasoft running 
of the spin-independent singlet potentials up to ${\cal O}(1/m^2)$, 
and the corresponding contribution to the spectrum. This includes 
the static energy at next-to-next-to-next-to-leading logarithmic (NNNLL) 
order. As a byproduct of these results we set the stage for the complete analytic and numerical computation of the heavy 
quarkonium spectrum with N$^3$LL accuracy for $l\not=0$ (angular momentum) 
and $s=0$ (spin) states. We also compute the next-to-next-to-next-to-next-to-leading order (N$^4$LO) ultrasoft spin-independent 
contribution to the heavy quarkonium mass and static energy.
\end{abstract}
\maketitle

\section{Introduction}
The evaluation of the three-loop soft contribution to the static potential 
\cite{Anzai:2009tm,Smirnov:2009fh} has given the final piece needed for the complete evaluation of the 
heavy quarkonium spectrum at NNNLO \cite{Kniehl:2002br,Penin:2002zv}. This result is also necessary 
for the long term project of obtaining the heavy quarkonium spectrum with 
NNNLL accuracy\footnote{Actually, this accuracy has already been achieved for the hyperfine splitting \cite{Kniehl:2003ap,Penin:2004xi}.}, 
and the non-relativistic sum rules and 
$t$-$\bar t$ production near threshold with NNLL/NNNLO one. It is then timely 
to compute the next-to-leading ultrasoft running of the potentials, as 
it enters in the evaluation of the previously mentioned observables with such precision. 
Therefore, in this paper we compute the next-to-leading ultrasoft running of 
the spin-independent potentials. For the static potential the running has already been 
computed in Ref. \cite{Brambilla:2009bi}, we confirm this result, for the $1/m$ and $1/m^2$ potentials the result is new. 

We will write the results for the potentials in position space. This could be 
eventually convenient for future numerical evaluations of decays and sum rules. 
Moreover, such computation is interesting on its own, as it shows the subtleties appearing 
in the explicit matching (and the associated scheme dependence) 
between the soft and ultrasoft computation. In momentum space this discussion has already been 
made in Ref. \cite{Kniehl:2002br}, we do so in position space (see also 
Ref. \cite{Pineda:2010mb} for a similar discussion for the QCD static potential in three dimensions, 
 and Ref. \cite{Pineda:2007kz} for the SUSY QCD static potential). This
discussion is relevant, as the subtraction scheme of the ultrasoft divergences could be 
different in position and momentum space. 

The basis of potentials that we use for the $1/m$ and $1/m^2$ potentials 
is redundant. Therefore, 
the expression obtained for each potential is ambiguous, since field redefinitions can shift 
some contributions from the $1/m$ to the $1/m^2$ potentials, and viceversa. In principle, one could avoid this problem by considering all 
the $1/m$ and $1/m^2$ potentials as a whole. Yet, this observation has implications in one 
of the possible applications of our result: comparison (at short distances) with the recent lattice simulations of the $1/m$ and $1/m^2$  
potentials obtained in Refs. \cite{Koma:2006si,Koma:2006fw}
using their non-perturbative 
expression in terms of Wilson loops \cite{Brambilla:2000gk,Pineda:2000sz}.  
Unfortunately, the ambiguity just mentioned makes not possible a direct connection 
between the non-perturbative expressions of the potentials in terms of Wilson 
loops and the perturbative computation. This would require a dedicated study that 
goes beyond the aim of this work.

Whereas there are still some pieces left for a complete NNNLL evaluation of the 
heavy quarkonium spectrum, our result provides the missing link for the complete 
result for  $l\not=0$ and $s=0$ states, the structure of 
which is given in this paper for the first time. The full explicit analytic form will be presented elsewhere.
Another by-product of our computation is the  N$^4$LO ultrasoft spin-independent 
contribution to the heavy quarkonium mass.

The next-to-leading ultrasoft running of the $1/m^2$ potential was computed in Ref. \cite{Hoang:2006ht} in a different framework named vNRQCD (see Refs. \cite{Luke:1999kz,Manohar:1999xd,Hoang:2002yy}). The computation of such object alone does not make much sense due to field redefinitions ambiguities that can shift contributions among different potentials. If we compare expressions for the $1/m^2$ potential, we disagree with the running of $V^{(2)}_{\bf p^2}$ obtained in that paper but yet, as mentioned, this could be due to field redefinition ambiguities. In Ref. \cite{Max2} the $1/m^2$ and $1/m$ potentials were considered. Nevertheless, no definite outcome for the complete NLL ultrasoft running of the potentials was obtained. The reason was that the result was dependent on how the infrared divergences were regulated, so they even ended up having more than one possible result. Thus, a complete comparison with their results at this stage is not possible. Other issues that complicate the comparison are that  the soft contribution is not included in their potential and the intrinsic scheme dependence. Either way, we believe that such future comparison should be better performed for specific observables. In this respect a future evaluation of the heavy quarkonium spectrum with NNNLL accuracy for $l\not=0$ and $s=0$ states with vNRQCD would be a good object for such comparison.

Non-perturbative effects will not be considered in this paper and ultrasoft effects will be computed within perturbation theory (i.e. in the $m\al^2 \gg \lQ$ limit), yet the renormalization group (RG) results will also be valid when $m\al^2 \sim \lQ$. The ultrasoft effects can be easily obtained in the 
non-equal mass case, as they only depend on the reduced mass. On the other hand the full soft contribution is 
only known in the equal mass case at the appropriate order. 

The outline of the paper is as follows.
In Sec.~\ref{pNRQCD} we introduce the theoretical setup and show some bare results relevant for our 
computation.  In Sec.~\ref{sec:deltaEus} we compute the ultrasoft correction to 
the spin-independent Hamiltonian with NLL precision and give explicit expressions for the relevant potentials. 
 In Sec.~\ref{Observables} we give expressions for the heavy quarkonium energy and static potential with NNNLL precision 
(for the heavy quarkonium mass only when $l\not=0$ and $s=0$),  
and in Sec.~\ref{conclusion} we present our conclusions. Finally, in the Appendix we gather some constants that 
appear throughout the computation.

\section{pNRQCD}
\label{pNRQCD}

Up to NLO in the multipole expansion
the effective Lagrangian density of pNRQCD 
 takes the form~\cite{Pineda:1997bj,Brambilla:1999xf}:
\begin{eqnarray}
& & {\cal L}_{\rm us} =
{\rm Tr} \Biggl\{ {\rm S}^\dagger \left( i\partial_0  - h_s(r)   \right) {\rm S} 
 + {\rm O}^\dagger \left( iD_0 - h_o(r)  \right) {\rm O} \Biggr\} 
\nonumber\\
& &\qquad
+ g V_A ( r) {\rm Tr} \left\{  {\rm O}^\dagger {\bf r} \cdot {\bf E} \,{\rm S}
+ {\rm S}^\dagger {\bf r} \cdot {\bf E} \,{\rm O} \right\} 
  + g {V_B (r) \over 2} {\rm Tr} \left\{  {\rm O}^\dagger \left\{{\bf r} \cdot {\bf E} , {\rm O}\right\}\right\} .
\label{pnrqcd0}
\end{eqnarray}
We define color singlet and octet fields for the quark-antiquark system by $S = S({\bf r},{\bf R},t)$ and 
$O^a =  O^a({\bf r},{\bf R},t)$, respectively. ${\bf R} \equiv \frac{m_1}{m_1+m_2}{\bf x}_1+\frac{m_2}{m_1+m_2}{\bf x}_2$ 
is the center position of the system, and ${\bf r}={\bf x}_1-{\bf x}_2$.
In order for $S$ and $O^a$ to have the proper free-field normalization in color space they are related to the fields in Eq.~(\ref{pnrqcd0}) as follows:
\begin{equation}
{\rm S} \equiv { 1\!\!{\rm l}_c \over \sqrt{N_c}} S\,, \qquad {\rm O} \equiv  { T^a \over \sqrt{T_F}}O^a. 
\label{norm}
\end{equation}
All gluon and scalar fields in Eq. (\ref {pnrqcd0}) are evaluated 
in ${\bf R}$ and the time $t$, in particular the chromoelectric field ${\bf E} \equiv {\bf E}({\bf R},t)$ and the ultrasoft covariant derivative
$iD_0 {\rm O} \equiv i \partial_0 {\rm O} - g [A_0({\bf R},t),{\rm O}]$. 

 $h_s$ can be split in the kinetic term and the potential: 
\bea
& &
h_s({\bf r}, {\bf p}, {\bf S}_1,{\bf S}_2) = 
{{\bf p}^2 \over 2\, m_r}+ 
V_s({\bf r}, {\bf p}, {\bf S}_1,{\bf S}_2), 
\\
& & 
h_o({\bf r}, {\bf p}, {\bf S}_1,{\bf S}_2) = 
 {{\bf p}^2 \over 2\, m_r}  
 + 
V_o({\bf r}, {\bf p}, {\bf S}_1,{\bf S}_2),
\eea
where $m_r=m_1m_2/(m_1+m_2)$, ${\bf p}=-i\bfnabla_{\bf r}$ and 
${\bf S}_1$ and ${\bf S}_2$ are the spin of the quark and the antiquark 
respectively. 
For the equal mass case: $m_1=m_2=m$, the potential has the following structure 
(we drop the labels $s$ and $o$ for the singlet and octet, which have to be understood):
\bea
V(r)&=&V^{(0)}(r) +{V^{(1)}(r) \over m}+{V^{(2)} \over m^2}+\cdots,  
\label{ppot}                                                                      
\\
V^{(2)}&=&V^{(2)}_{SD}+V^{(2)}_{SI},\nn\\
V^{(2)}_{SI}
&=&                                   
{1 \over 2}\left\{{\bf p}^2,V_{{\bf p}^2}^{(2)}(r)\right\}
+{V_{{\bf L}^2}^{(2)}(r)\over r^3}{\bf L}^2 + V_r^{(2)}(r),
\\
\label{VSD}
V^{(2)}_{SD} &=&
V_{LS}^{(2)}(r){\bf L}\cdot{\bf S} + V_{S^2}^{(2)}(r){\bf S}^2
 + V_{{\bf S}_{12}}^{(2)}(r){\bf S}_{12}({\hat {\bf r}}), 
\eea
where ${\bf S}={\bf S}_1+{\bf S}_2$,  ${\bf S}_{12}({\hat {\bf r}}) \equiv 
4(3 {\hat {\bf r}}\cdot {\bf S}_1 \,{\hat {\bf r}}\cdot {\bf S}_2 - {\bf S}_1\cdot {\bf S}_2)$, and ${\bf L}={\bf r}\times {\bf p}$. The 
last two equalities hold in 4 dimensions. ${\bf L}^2$ is generalized to (${\bf q}={\bf p}-{\bf p}'$)
\begin{equation}
\frac{{\bf L}^2}{2\pi r^3} \to 
\left(\frac{{\bf p}^2-{\bf p}^{\prime\,2}}{{\bf q}^2}\right)^2 - 1
\label{lddef}
\end{equation}
to be compatible with $D$ dimensional calculations in momentum
space. 
In this paper we focus on the ultrasoft corrections to the spin-independent singlet potentials. Therefore, we will not consider Eq. (\ref{VSD}) (nor 
its generalization to $D$ dimensions, see for instance Ref. \cite{Pineda:2006ri}) in the following and drop the labels $s$ and $o$, except for the static potentials. 

From now on we will use the index ``$B$'' to explicitly denote bare quantities. Parameters without this index are understood to be renormalized. 
The bare parameters of the theory are $\al_B$ ($g_B$) and the potentials, generically denoted by $V_B$. $\al$ and $V_{\{s,o,A,B\}}(r)$ are the
 Wilson coefficients of the effective Lagrangian. They are fixed 
 at a scale $\nu$ smaller than (or similar to) $1/r$ and larger than the ultrasoft and any other scale in the problem by 
 matching pNRQCD and NRQCD.

In our convention 
$\al_B$ is dimensionless and related to $g_B$ by ($D=d+1=4+2\epsilon$)
\be
\alpha_B=\frac{g_B^2\nu^{2\epsilon}}{4\pi}\,,
\ee
where $\nu$ is the renormalization scale.
It has a special status since 
it does not receive corrections from other Wilson coefficients of the 
effective theory. We renormalize it multiplicatively:
\be
\al_B=Z_{\al}\al
\,,
\ee
where
\be
Z_{\al}
=1+
\sum_{s=1}^{\infty}Z^{(s)}_{\al}\frac{1}{\epsilon^s}\,.
\ee
The RG equation of $\al$ is
\be
\nu\frac{d}{d\nu}\al\equiv \al\beta(\al;\epsilon)=2\epsilon\al+\al\beta(\al;0)\,.
\ee
In the limit $\epsilon \rightarrow 0$
\be
\nu\frac{d}{d\nu}\al\equiv \al\beta(\al;0)
\equiv \al\beta(\al)=-2\al \frac{d}{d\al}Z^{(1)}_{\al}\,,
\ee
where
\be
Z^{(1)}_{\al}=\frac{\al}{4\pi}\beta_0+\cdots 
\qquad 
\al\beta(\al)=-2\al\left(\beta_0\frac{\al}{4\pi}+\beta_1\frac{\al^2}{(4\pi)^2}
+\cdots 
\right) 
\,,
\ee
and expressions for $\beta_0$, $\beta_1$, etc, can be found in the Appendix. 

The bare potentials $V_B$ in position space have integer mass dimensions (note that this is not so in 
momentum space) and, due to the structure of the theory, we do not renormalize them multiplicatively
(see the discussion in Ref.~\cite{Pineda:2000gz}). We define
\be
\label{VBsplitting}
V_B=V+\delta V\,.
\ee
$\delta V$ will generally depend on the (matching) coefficients of the effective theory, i.e. on $\al$ and $V$, 
and on the number of space-time dimensions.
In $D$ dimensions, using the MS renormalization scheme, we define
\be
\delta V
=
\sum_{s=1}^{\infty}Z^{(s)}_{V}\frac{1}{\epsilon^s}\,.
\ee

From the scale independence of the bare potentials
\be
\nu \frac{d}{d\nu}V_B=0
\,,
\ee
one obtains the RG equations of the different
renormalized potentials. They can schematically be written as one (vector-like) equation including all potentials:
\be
\nu \frac{d}{d\nu}V=B(V)\,, \label{VRGE}
\ee
\be
B(V)\equiv -\left(\nu \frac{d}{d\nu}\delta V\right)\,.
\ee
Note that Eq.~(\ref{VRGE}) implies that all the $1/\epsilon$ poles disappear once the derivative 
with respect to the renormalization scale is performed.
 This imposes some constraints on $\delta V$:
\bea
\label{Z1}
{\cal O}(1/\epsilon): \qquad
&&B(V)=-2\al \frac{\partial}{\partial\al}Z^{(1)}_{V}\,,\; \label{1loopBV}
\\
{\cal O}(1/\epsilon^2): \qquad
&&
\label{Z2}
B(V)\frac{\partial}{\partial V}Z^{(1)}_{V}
+
\al \beta (\al)\frac{\partial}{\partial\al}Z^{(1)}_{V}
+
2\al \frac{\partial}{\partial\al}Z^{(2)}_{V}=0\,,
\eea
and so on. 

Out of this theory we can obtain some observables. In this paper we focus on the energy of the heavy quarkonium (also in its static limit). The singlet propagator near on-shell can be approximated to the following expression
\bea
\nn
\int dt e^{iEt}d^3R
\langle vac |S(t,{\bf r},{\bf R})S^{\dagger}(0,{\bf r}',{\bf 0})|vac \rangle 
&\sim& \phi_n({\bf r})\phi_n({\bf r}')
\langle n|
\frac{i}{E-h_s^B-\Sigma_B(E)+i\epsilon}|n\rangle 
\\
&\sim& \phi_n({\bf r})\phi_n({\bf r}')\frac{i}{E-E_n^{pot}-\delta E_n^{us}+i\epsilon}
\,,
\eea
where $n$ generically denotes the quantum number of the bound state: 
$n \rightarrow$ ($n$ (principal quantum number), $l$ (orbital angular momentum), $s$ (total spin), $j$ (total angular momentum)). 
$E_n^{pot}$ and $\phi_n({\bf r})$ are the eigenvalue and eigenfunction respectively of the equation
\be
h_s\phi_n({\bf r})=E_n^{pot}\phi_n({\bf r})
\ee
and, in general, will depend on the renormalization scheme the ultrasoft computation has been performed with.
The self-energy $\Sigma_B(E)$ accounts for the effects due to the 
ultrasoft scale and can be expressed in a compact form at NLO in the multipole expansion (but exact to any order in $\al$) through the 
chromoelectric correlator. It reads (in the Euclidean)
\be
\Sigma_{B}(E)
  = V_A^2{T_F \over (D-1) N_c}  \int_0^\infty \!\! dt {\bf r} e^{-t(h^B_{o}-E)} {\bf r}
  \langle vac|
g{\bf E}_E^a(t) 
\phi_{\rm adj}^{ab}(t,0) g{\bf E}_E^b(0) |vac \rangle
\label{deltaVUS}
\,.
\end{equation}
The pNRQCD one-loop computation yields \cite{Pineda:1997ie,Brambilla:1999qa,Kniehl:1999ud}
\be
\Sigma_{B}({\rm 1-loop})
=
-g^2C_fV_A^2(1+\epsilon)\frac{\Gamma(2+\epsilon)\Gamma(-3-2\epsilon)}{\pi^{2+\epsilon}}
{\bf r}\,(h^B_{o}-E)^{3+2\epsilon}{\bf r}\,.
\label{USbare1loop}
\ee
The two-loop bare expression can be trivially deduced from 
the results obtained in Refs.~\cite{Eidemuller:1997bb,Brambilla:2006wp,MaxHyb} for the static case. It reads
\be
\label{USbare2loop}
\Sigma_{B}({\rm 2-loop})
=
g^4C_fC_AV_A^2\Gamma(-3-4\epsilon)
\left[
{\cal D}^{(1)}(\epsilon)-(1+2\epsilon){\cal D}_1^{(1)}(\epsilon)
\right]
{\bf r}\,(h^B_{o}-E)^{3+4\epsilon}{\bf r}\,,
\ee
where
\be
{\cal D}^{(1)}(\epsilon)
=
\frac{1}{(2\pi)^2}\frac{1}{4\pi^{2+2\epsilon}}\Gamma^2(1+\epsilon)g(\epsilon)\,,
\ee
\be
{\cal D}_1^{(1)}(\epsilon)
=
\frac{1}{(2\pi)^2}\frac{1}{4\pi^{2+2\epsilon}}\Gamma^2(1+\epsilon)g_1(\epsilon)\,,
\ee
and
\be
g(\epsilon)=\frac{2 \epsilon^3+6 \epsilon^2+8 \epsilon+3}{\epsilon \left(2 \epsilon^2+5 \epsilon+3\right)}
-\frac{2 \epsilon \Gamma
   (-2 \epsilon-2) \Gamma (-2 \epsilon-1)}{(2 \epsilon+3) \Gamma (-4 \epsilon-3)}
\,,
\ee
\be
g_1(\epsilon)=\frac{6 \epsilon^3+17 \epsilon^2+18 \epsilon+6}{\epsilon^2 
\left(2 \epsilon^2+5 \epsilon+3\right)}+\frac{4 (\epsilon+1)
   n_f T_F}{\epsilon (2 \epsilon+3) N_c}+\frac{2
   \left(\epsilon^2+\epsilon+1\right) \Gamma (-2 \epsilon-2) \Gamma (-2 \epsilon-1)}{\epsilon (2 \epsilon+3)
   \Gamma (-4 \epsilon-3)}
\,.
\ee
From $\Sigma_B(E)$ it is possible to obtain $\delta E_n^{us}$. This will be discussed in the following sections.

In principle we should also consider possible soft and ultrasoft corrections to $V_A$. Those have been studied in Ref. 
\cite{Pineda:2000gz} with LL accuracy, in Ref. \cite{Brambilla:2006wp} with NLO accuracy, and in Ref. \cite{Brambilla:2009bi} with NLL accuracy,
reaching to the conclusion that they do not contribute to the precision of our computation (so we can set
$V_A=1$):
\be
\nu\frac{d}{d\nu}V_A=0+{\cal O}(\al^3)
\,,
\ee
whereas for the initial matching condition $V_A=1+{\cal O}(\al^2)$. 

\section{$V_s$}
\label{sec:deltaEus}

\subsection{Ultrasoft running}
We now discuss how to obtain the ultrasoft RG running of $V_s$ from $\Sigma_B(E)$. 
The bare ultrasoft 
self-energy is a function of $h_o-E$, $\al$ and $\epsilon$, where this $\al$ refers to the one associated to the ultrasoft running. The $\al$ associated to the soft 
running is encoded in the matching coefficients (i.e. the potentials).
From the combined one and two loop ultrasoft computation we obtain 
(after changing the bare alpha by the renormalized one) 
\bea
\nn
\Sigma_{B}(E)
&=&
-
\frac{1}{\epsilon^2}C_fV_A^2{\bf r}(h_o-E)^3{\bf r} 
\frac{2}{3}\beta_0\frac{\al^2(\nu)}{(4\pi)^2}
\\
&&
\nn
-\frac{1}{\epsilon}C_fV_A^2
{\bf r}(h_o-E)^3{\bf r} 
\left[
\frac{\al(\nu)}{3\pi}
-
\frac{\al^2(\nu)}{36\pi^2}
(C_A(-\frac{47}{3}-2\pi^2)+\frac{10}{3}T_Fn_f)
\right]
\\
&&
\nn
+C_f V_A^2{\bf r}(h_o-E)^3 
\left[
-\frac{\al(\nu) }{9 \pi } \left(6 \ln \left(\frac{h_o-E}{\nu }\right)+6\ln 2-5\right)
\right.
\\
&& 
\nn  
+\frac{\al^2(\nu) }{108 \pi ^2} 
\left(18 \beta_0 \ln ^2\left(\frac{h_o-E}{\nu
   }\right)-6 \left(C_A \left(13+4 \pi ^2\right)+2 \beta_0 (5-3\ln 2)\right) \ln \left(\frac{h_o-E}{\nu
   }\right)
   \right.
   \\
   &&
   \nn
-2 C_A \left(-84+39 \ln 2+4 \pi ^2 (-2+3\ln 2)+72 \zeta (3)\right)
\\
&&
\left.
+\beta_0 \left(67+3 \pi ^2-60 \ln 2+18\ln 2\right)\right)
   \Bigg]{\bf r}
+{\cal O}(\epsilon,\al^3)
\,.
\eea
Note that the $1/\epsilon^2$ term comes from two sources: the two loop bare result and the $1/\epsilon$ inside $\al_B$ in the one loop ultrasoft.
Quite remarkably, there is no log dependence of the $1/\epsilon$ and $1/\epsilon^2$ terms. This is of 
fundamental importance for renormalizability and a check of consistency. 
This result has to be reexpressed in terms of the potentials of the 
singlet/octet Hamiltonian and $h_s-E$. Positive powers of $h_s-E$ do not contribute to the energy, as they cancel powers of $1/(h_s-E)$ in the Green function. Therefore, we are rather interested in the identity (valid in $D$ dimensions)
\bea
\nn
&&
{\bf r}(h_o-E)^3{\bf r}=
{\bf r}^2(\Delta V)^3
-\frac{1}{2m_r^2}\left[{\bf p},\left[{\bf p},V^{(0)}_o\right]\right]
+\frac{1}{2m_r^2}\left\{{\bf p}^2,\Delta V\right\}
+\frac{2}{m_r}\Delta V  \left(r\frac{d}{dr}V^{(0)}_s\right)
\\
&&
\nn
\qquad\qquad
+\frac{1}{2m_r}\left[
(\Delta V)^2(3d-5)+4\Delta V\left( \left(r\frac{d}{dr}\Delta V\right)+\Delta V \right)
+\left( \left(r\frac{d}{dr}\Delta V\right)+\Delta V \right)^2
\right]
\\
&&
\qquad\qquad
+{\cal O}((h_s-E))
\,,
\eea
where we have approximated $h_o-h_s=V^{(0)}_o-V_s^{(0)}$, which is enough for our precision, and
defined $\Delta V \equiv V^{(0)}_o-V^{(0)}_s$.
We have used the combination $\left( \left(r\frac{d}{dr}\Delta V\right)+\Delta V \right)$,
since it has a ${\cal O}(\epsilon)$ suppression with respect to $\Delta V$. Note that in $D$ dimensions 
the static potential has the following expansion in terms of the bare coupling constant ($C_s\equiv -C_F$ and $C_o \equiv 1/(2N_c)$):
\bea
\label{VsBD}
V^{(0)}_{s/o,B}
&=&
C_{s/o}\,g_B^2\sum_{n=0}^{\infty}\frac{g_B^{2n}c^{(s/o)}_n(D)r^{-2(n+1)\epsilon}}{r}\,.
\eea

We can now obtain the counterterms of the singlet Hamiltonian due 
to the ultrasoft divergences up to NLO in the following compact expression
\bea
\nn
\delta V_s
&=&
\Biggl(
{\bf r}^2(\Delta V)^3
-\frac{1}{2m_r^2}\left[{\bf p},\left[{\bf p},V^{(0)}_o\right]\right]
+\frac{1}{2m_r^2}\left\{{\bf p}^2,\Delta V\right\}
+\frac{2}{m_r}\Delta V  \left(r\frac{d}{dr}V^{(0)}_s\right)
\\
&&
\nn
+\frac{1}{2m_r}\left[
(\Delta V)^2(3d-5)+4\Delta V\left( \left(r\frac{d}{dr}\Delta V\right)+\Delta V \right)
+\left( \left(r\frac{d}{dr}\Delta V\right)+\Delta V \right)^2
\right]
\Biggr)
\\
&&
\times
\Biggl[
\frac{1}{\epsilon}
C_fV_A^2
\left[
\frac{\al(\nu)}{3\pi}
-
\frac{\al^2(\nu)}{36\pi^2}
(C_A(-\frac{47}{3}-2\pi^2)+\frac{10}{3}T_Fn_f)
\right]
+
\frac{1}{\epsilon^2} C_fV_A^2
\frac{2}{3}\beta_0\frac{\al^2(\nu)}{(4\pi)^2}
\Biggr]
\,.
\label{deltaVs}
\eea
The whole $1/\epsilon^2$ term fulfills Eq. (\ref{Z2}) and hence it is a check of the two-loop computation. 
In the above expressions we choose to keep the full $d$-dependence (also in the potentials, see Eq. (\ref{VsBD})). This 
is not particularly relevant for the 
computations we perform in this paper, but will be potentially important once potential divergences are included, necessary for a 
complete NNNLL evaluation of the heavy quarkonium mass. As we have already mentioned,
 in this prescription we also take $V_s^{(0)}$ and $V^{(0)}_o$ in $d$ dimensions. Since we only need them at one loop, 
 the ultrasoft  divergences of $V^{(0)}_{s/o}$ do not show up yet. This actually means that the renormalized potentials are 
equal to their bare expressions (which one can find in Ref. \cite{Schroder:1999sg}): $V_{s/o}^{(0)}\simeq V_{s/o,B}^{(0)}$, and 
$V^{(0)}_{s/o,B}$ are finite when we take the $\epsilon \rightarrow 0$ limit. Moreover, at one loop, we also have the equality $V_o^{(0)}=-1/(N_c^2-1)V_s^{(0)}$.
We prefer the scheme defined by Eq. (\ref{deltaVs}), since it allows to keep the ultrasoft counterterms in a very compact manner. One is always free 
to change to a more standard $\MS$ scheme. Note that the $\MS$ scheme in momentum and position space are different. 

Using Eq. (\ref{deltaVs}) and Eq. (\ref{Z1}) we obtain the following RG equation that resums the ultrasoft logs of the potential: 
\be
\nu\frac{d}{d\nu} V_{s,\MS}
=
B_{V_s}
\,,
\ee 
where
\bea
B_{V_s}&=&
C_fV_A^2
\left[
{\bf r}^2(\Delta V)^3
+\frac{2}{m_r}\Biggl(\Delta V  \left(r\frac{d}{dr}V^{(0)}_s\right)
+
(\Delta V)^2
\Biggr)
-\frac{1}{2m_r^2}\left[{\bf p},\left[{\bf p},V^{(0)}_o \right]\right]
\right.
\\
\nn
&&
\left.
+\frac{1}{2m_r^2}\left\{{\bf p}^2,\Delta V \right\}
\right]
\times
\left[
-\frac{2\al(\nu)}{3\pi}
+
\frac{\al^2(\nu)}{9\pi^2}
(
C_A(-\frac{47}{3}-2\pi^2)+\frac{10}{3}T_Fn_f
)
+
{\cal O}(\al^3)
\right]
\,,
\eea
and now one could take the four-dimensional expression for the potentials. This result holds true in both schemes, the MS and $\MS$ (in a way this is due to the 
fact that the subdivergencies associated to $\al$ also change to make the result scheme independent). 
After solving the RG equation we find
\bea
\nn
\delta V_{s,RG}(r;\nu_s,\nu)
&=&
\left[
\Biggl(
{\bf r}^2(\Delta V)^3
+\frac{2}{m_r}\left(\Delta V  \left(r\frac{d}{dr}V^{(0)}_s\right)+(\Delta V)^2\right)
\Biggr)F(\nu_s;\nu)
\right.
\\
&&
\left.
-\frac{1}{2m_r^2}\left[{\bf p},\left[{\bf p},V^{(0)}_o(r) F(\nu_s;\nu)\right]\right]
+\frac{1}{2m_r^2}\left\{{\bf p}^2,\Delta V(r) F(\nu_s;\nu)\right\}
\right]
\label{VsRG}
\,,
\eea
where we define
\bea
F(\nu_s;\nu)
&=&C_fV_A^2
\frac{2\pi}{\beta_0}
\left\{
\frac{2}{3\pi}\ln\frac{\al(\nu)}{\al(\nu_s)}
\right.
\\
\nn
&&
\left.
-(\al(\nu)-\al(\nu_s))
\left(
\frac{8}{3}\frac{\beta_1}{\beta_0}\frac{1}{(4\pi)^2}-\frac{1}{27\pi^2}\left(C_A\left(47+6\pi^2\right)-10T_Fn_f\right)
\right)
\right\}
\\
\nn
&\simeq&
-C_FV_A^2\frac{2}{3}\frac{\al}{\pi}\ln\frac{\nu}{\nu_s}+{\cal O}(\al^2)
\,.
\eea
Note that, formally, we can set $\nu_s=1/r$ as far as $F(1/r,\nu)$ is kept inside the 
(anti)commutators, i.e. in the way displayed in Eq. (\ref{VsRG}). 
Yet one should be careful, as objects that are ill defined (not even distributions) could appear. We discuss this issue further below.

From Eq. (\ref{VsRG}) we can easily identify the RG contribution to each potential\footnote{And from previous equations one could also 
easily obtain the counterterms and RG equations for each potential.} (now we work in the equal mass case). For the static potential we have
\be
\delta V^{(0)}_{s,RG}(r;\nu_s,\nu)
=
{\bf r}^2(\Delta V)^3
F(\nu_s;\nu)
\,,
\ee
which agrees with Ref. \cite{Brambilla:2009bi}.

For the $1/m$ potential the RG contribution reads
\be
\delta V^{(1)}_{s,RG}(r;\nu_s,\nu)
=4
\Biggl(
\Delta V  \left(r\frac{d}{dr}V^{(0)}_s\right)
+
(\Delta V)^2
\Biggr)F(\nu_s;\nu)
\,.
\ee

For the momentum-dependent $1/m^2$ potential we have
\bea
\delta V^{(2)}_{p^2,RG}(r;\nu_s,\nu)
=
4 \Delta V(r) F(\nu_s;\nu)
\label{Vp2RG}
\,.
\eea
The RG correction to $V_r$ deserves a special discussion. From Eq. (\ref{VsRG}) we have 
\be
\delta V^{(2)}_{r,RG}(r;\nu_s,\nu)
=
-2\left[{\bf p},\left[{\bf p},V^{(0)}_o F(\nu_s;\nu)\right]\right]
=\int \frac{d^3q}{(2\pi)^3}e^{i{\bf q}\cdot {\bf r}}\delta \tilde V^{(2)}_{r,RG}(q;\nu_s,\nu)
\label{VsrRG}
\,,
\ee
where
\be
\delta \tilde V^{(2)}_{r,RG}(q;\nu_s,\nu)
=
-2 {\bf q}^2\tilde V^{(0)}_o(q)F(\nu_s;\nu)
\,,
\ee
and $\tilde V^{(0)}_o$ is the Fourier transform of $V^{(0)}_o$.
Note that we have chosen to write $\delta V_r$ in momentum space as well. This we could also do for the previous potentials but now will be
particularly convenient. The reason is that for $\delta V_{r,RG}(r;\nu_s,\nu)$ we cannot naively set $\nu_s=1/r$
(as we did for the previous potentials and was convenient to resum logs of $r$)
because divergent distributions like $\delta^{(3)}({\bf r})\ln r$ could appear. On the other hand we 
would like to make something analogous, as it allows to resum (and keep in a compact way) some subleading logs. We then choose to 
set $\nu_s=q$ in the Fourier transform and define
\be
\delta V^{(2)}_{r,RG}(r;\nu)
=
\int \frac{d^3q}{(2\pi)^3}e^{i{\bf q}\cdot {\bf r}}\delta \tilde V^{(2)}_{r,RG}(q;q,\nu)
\,,
\ee
where $\delta \tilde V^{(2)}_{r,RG}(q;q,\nu)$ is Taylor expanded in powers of $\ln q$.
We will elaborate on this expression in the following subsection (in practice we will only need the single log of this expansion).
Finally also note that $\delta V^{(2)}_{r,RG}$ vanishes in the large $N_c$ limit.

\subsection{Initial matching conditions}

The only thing left to obtain is the initial matching condition for the potential. 
We will only consider the spin-independent terms and stick to the equal mass case, since 
the matching coefficients at one loop are not known in the non-equal mass case, although many partial results exist.

We now study each potential separately. Note that the result will depend on the basis 
of potentials used and on field redefinitions except for the singlet static potential, which is 
unambiguous\footnote{This observation also applies to the RG corrections to the potentials obtained in 
the previous subsection.}.

The initial matching conditions for the static potential in our $\MS$ scheme reads 
\begin{eqnarray}
\label{V0sMS}
V^{(0)}_{s,\MS}(r;\nu)
&=&
 -\frac{C_f\,\alpha_s(\nu)}{r}\,
\bigg\{1+\sum_{n=1}^{3}\left(\frac{\alpha_s(\nu)}{4\pi}\right)^n a_n(\nu;r)\bigg\}
\,,
\end{eqnarray}
with coefficients
\begin{eqnarray}
a_1(\nu,r)
&=&
a_1+2\beta_0\,\ln\left(\nu e^{\gamma_E} r\right)
\,,
\nonumber\\
a_2(\nu,r)
&=&
a_2 + \frac{\pi^2}{3}\beta_0^{\,2}
+\left(\,4a_1\beta_0+2\beta_1 \right)\,\ln\left(\nu e^{\gamma_E} r\right)\,
+4\beta_0^{\,2}\,\ln^2\left(\nu e^{\gamma_E} r\right)\,
\,,
\nonumber \\
a_3(\nu,r)
&=&
a_3+ a_1\beta_0^{\,2} \pi^2+\frac{5\pi^2}{6}\beta_0\beta_1 +16\zeta_3\beta_0^{\,3}
\nonumber \\
&+&\bigg(2\pi^2\beta_0^{\,3} + 6a_2\beta_0+4a_1\beta_1+2\beta_2+\frac{16}{3}C_A^{\,3}\pi^2\bigg)\,
  \ln\left(\nu e^{\gamma_E} r\right)\,
\nonumber \\
&+&\bigg(12a_1\beta_0^{\,2}+10\beta_0\beta_1\bigg)\,
  \ln^2\left(\nu e^{\gamma_E} r\right)\,
+8\beta_0^{\,3}  \ln^3\left(\nu e^{\gamma_E} r\right)\,
\,.
\label{eq:Vr}
\end{eqnarray}
Explicit expression for $a_i$ can be found in the literature
\cite{FSP,Schroder:1999sg,Brambilla:1999qa,Kniehl:1999ud,Pineda:2000gz,Anzai:2009tm,Smirnov:2009fh}. 
For ease of reference we display them in the Appendix. 

We choose to write the initial matching conditions of the potentials, like Eq. (\ref{V0sMS}), in terms of the (single) factorization scale of the effective theory $\nu$ and not $\nu_s$, as $\nu_s$ does not appear in a matching computation done order by order in $\alpha$. The (left-over) factorization scale dependence of the matching coefficient would cancel with the scale dependence of loops in the effective theory (in practice most of the scale dependence in the potentials, for instance in Eq. 
(\ref{V0sMS}), cancels with the scale dependence of $\alpha(\nu)$, effectively becoming $\alpha(1/r)$ but this does not change the physical principle). When a RG analysis is done one introduces two factorization scales: the one where the running starts, which we have named $\nu_s$ in the previous section, and the factorization scale that would cancel with the scale dependence of loops in the effective theory, which we have always named $\nu$. Therefore, the total RG improved static potential then reads
\be
\label{VRGIsMS}
V^{(0),RG}_{s,\MS}(r;\nu)=V^{(0)}_{s,\MS}(r;\nu_s)+\delta V^{(0)}_{s,RG}(r;\nu_s,\nu)
\,,
\ee
and it is correct with NNNLL accuracy.

For the $1/m$ potential the initial matching condition reads
\be
V_{s,\MS}^{(1)}(r,\nu)
=
\frac{C_f\al^2(\nu)}{2r^2}
\left(
b_1+\frac{\al(\nu)}{\pi}\left[b_2+\left(\frac{b_1\beta_0}{2}-\frac{2}{3}(C_A^2+2C_AC_f)\right)\ln(\nu^2 r^2e^{2\gamma_E})\right]
\right)
\,,
\ee
where
\be
b_1=\frac{C_f}{2}-C_A\,,
\qquad
b_2=-\frac{89}{36}C_A^2+\frac{17}{18}C_AC_f+\frac{49}{36}C_AT_Fn_f-\frac{2}{9}C_fT_Fn_f
\,.
\ee
The two loop result has been taken from Ref. \cite{Kniehl:2001ju} and changed accordingly to fit our renormalization scheme for the 
ultrasoft computation. This explains the different coefficient $b_2$ we have compared with that reference. Ours is the proper 
one to be combined with the ultrasoft correction obtained in Eq. (\ref{deltaEusMSnl}) in the next section.
The total RG improved $1/m$ potential then reads
\be
\label{VRGI1MS}
V^{(1),RG}_{s,\MS}(r;\nu)=V^{(1)}_{s,\MS}(r;\nu_s)+\delta V^{(1)}_{s,RG}(r;\nu_s,\nu)
\,,
\ee
and it is correct with NNLL accuracy.

For the momentum-dependent $1/m^2$ potential the matching coefficient reads at one loop \cite{Kniehl:2002br}
\be
V_{p^2,\MS}^{(2)}(r,\nu) 
=
\frac{ C_f\al(\nu)}{4}
\left(
-4+\frac{\al(\nu)}{\pi}\left(-\frac{31}{9}C_A+\frac{20}{9}T_Fn_f-\left(\beta_0+\frac{8}{3}C_A\right)\ln(\nu^2 r^2e^{2\gamma_E})
\right)
\right)\,.
\ee
The total RG improved momentum-dependent $1/m^2$ potential then reads
\be
\label{VRGp21MS}
V^{(2),RG}_{p^2,\MS}(r;\nu)=V^{(2)}_{p^2,\MS}(r;\nu_s)+\delta V^{(2)}_{p^2,RG}(r;\nu_s,\nu)
\,,
\ee
and it is correct with NLL accuracy.

The results obtained for the above potentials are exact to the required accuracy, since their soft running is trivial and there are not 
potential loops at this order. This can be traced back to the fact that there is no dependence on the matching coefficients inherited from NRQCD, 
and no logarithms proportional to the mass appear. The independence of the potential on $\nu_s$, $\nu_s\frac{d V}{d \nu_s}=0$, can be easily implemented 
by setting $\nu_s=1/r$, giving information on the $\ln r$ dependence. In fixed-order bound state computations is 
convenient to work using $\al(\nu_s)$ as the expansion parameter with $\nu_s \sim m\al$. By expanding $\al(1/r)$ in powers of $\al(\nu_s)$ times $\ln(\nu_s r)$ subleading corrections appear that have to included with the appropriated precision.

The case of $V^{(2)}_r$ is different. It depends on the mass of the heavy quark (i.e. on the matching coefficients 
inherited from NRQCD). In this case we can not give the complete expression with NLL accuracy. For such accuracy we would need 
the soft divergences to a higher order and the inclusion of effects due to potential divergences. The latter are absorbed in 
delta-type potentials. Note that when computing their running at NLL the ultrasoft LL running would enter indirectly. 
Such kind of computations have already been undertaken in Refs. \cite{Kniehl:2003ap,Penin:2004xi} for the spin-dependent corrections to the 
hyperfine splitting of heavy quarkonium, a similar analysis for the spin-independent corrections would go much beyond the aim of this 
work. Either way we can still give the NLO matching coefficient for the potential, the relevant starting point for a complete NLL. 
It is first convenient to write it 
in terms of the potential in momentum space
\be
\label{Vrmatching}
V_{r,\MS}^{(2)}(r,\nu)
=
\int \frac{d^3q}{(2\pi)^3}e^{i{\bf q}\cdot {\bf r}}
\tilde V_{r,\MS}^{(2)}(q,\nu)
\,,
\ee
where
\bea
\tilde V_{r,\MS}^{(2)}(q,\nu)
&=&\pi C_f
\Bigg[
\al(q)(1+c_D(\nu)-2c_F^2(\nu))
\\
\nn
&&
\left.
+\frac{1}{\pi}(d_{vs}(\nu)+3d_{vv}(\nu)+\frac{1}{C_f}(d_{ss}(\nu)+3d_{vv}(\nu)))
 +\delta \tilde V_{soft}(\nu,q)
\right]
\,,
\eea
\be
\delta \tilde V_{soft}=\frac{\al^2}{\pi}
\left[
\left(\frac{9}{4}+\frac{25}{6}\ln\frac{\nu^2}{q^2}\right)C_A
+
\left(\frac{1}{3}-\frac{7}{3}\ln\frac{\nu^2}{q^2}\right)C_f
\right]
\,.
\ee
The soft one-loop result $\delta \tilde V_{soft}$ 
is taken from Ref. \cite{Kniehl:2002br}. The rest corresponds to the hard contribution.
We have checked that upon expanding the NRQCD matching coefficients in powers of 
$\al(\nu_s)$ we agree with that reference at ${\cal O}(\al^2)$. 
In order to use $\delta \tilde V_{soft}$ 
we had to change the scheme used for the Pauli ${\bfsigma}$ matrices in the computation of $d_{vv}$ in Ref. \cite{Pineda:1998kj}
 (see also the discussion in Refs. \cite{Manohar:2000hj,Pineda:2000sz}). The new expression for $d_{vv}$ can be found in the 
Appendix, as well as for the other NRQCD matching coefficients.
Eq. (\ref{Vrmatching}) is correct at LO, LL (soft). It is also correct at NLO, providing with the right initial matching condition to 
compute the NLL result. We can then give the following expression for the potential
\be
V_{r,\MS}^{(2),RG}(r;\nu)=V_{r,\MS}^{(2)}(r;\nu_s)+\delta V^{(2)}_{r,RG}(r;\nu_s,\nu)
\,.
\ee
This expression does not incorporate the NLL running associated to the Fourier transform of $\ln q$ yet. In order to do so it is rather more convenient 
to consider
\be
\label{Vr1}
V_{r,\MS}^{(2),RG}(r;\nu)=V_{r,\MS}^{(2)}(r)+\delta V^{(2)}_{r,RG}(r;\nu)
\,,
\ee
where
\be
V_{r,\MS}^{(2)}(r)
=
\int \frac{d^3q}{(2\pi)^3}e^{i{\bf q}\cdot {\bf r}}
\tilde V_{r,\MS}^{(2)}(q,q)
\,.
\ee
Although this new expression, Eq. (\ref{Vr1}), does not account for all the logs at NLL order, yet it does for the ultrasoft logs 
and those associated to $\ln q$. This is enough for the computation of the mass of the $l\not=0$ and $s=0$ states with NNNLL accuracy. The reason is that the delta potential 
does not contribute to the mass of those states. This is more clearly seen if we Taylor expand the above expression in powers of (the Fourier 
transform of) $\ln q$:
\bea
\label{Vrexpanded}
V_{r,\MS}^{(2),RG}(r;\nu)
&=&
\delta^3({\bf r})
\left(
\tilde V_{r,\MS}^{(2)}(\nu_s,\nu_s)+\delta \tilde V^{(2)}_{r,RG}(\nu_s;\nu_s,\nu)
\right.
\nn
\\
&&\qquad
\left.
-(\ln \nu_s) 
q\frac{d}{dq}(\tilde V_{r,\MS}^{(2)}(q,q) +\delta \tilde V^{(2)}_{r,RG}(q;q,\nu))|_{q=\nu_s} 
\right)
\nn
\\
&&
-\frac{1}{4\pi}
\left({\rm reg}\frac{1}{r^3}\right)q\frac{d}{dq}(\tilde V_{r,\MS}^{(2)}(q,q)+
\delta \tilde V^{(2)}_{r,RG}(q;q,\nu))|_{q=\nu_s}+\cdots
\,,
\eea
where we have used
\be
-\frac{1}{4\pi}
\left({\rm reg}\frac{1}{r^3}\right)
\equiv
\int \frac{d^3q}{(2\pi)^3}e^{i{\bf q}\cdot {\bf r}}
\ln q
\,.
\ee
Only the term proportional to ${\rm reg}\frac{1}{r^3}$ contributes to the mass of the states with $l\not=0$. Higher order terms in the Taylor 
expansion are subleading. The derivative of the potential can be basically written in terms of the NRQCD matching coefficients
\bea
q\frac{d}{dq}\tilde V_{r,\MS}^{(2)}(q,q)|_{q=\nu_s}
&=&C_f \al^2(\nu_s)
\left[
\frac{2}{3}T_Fn_f(c_D+c_1^{hl})+(\beta_0-\frac{13}{3}C_A)c_F^2
\right.
\nn
\\
&&
\left.
-\frac{\beta_0}{2}
+(\frac{14}{3}C_f-\frac{2}{3}C_A)c_k^2
\right]
,
\eea
\bea
q\frac{d}{dq}(\delta \tilde V^{(2)}_{r,RG}(q;q,\nu))|_{q=\nu_s}
=-C_f\al^2(\nu_s)\frac{16}{3}(\frac{C_A}{2}-C_f)
\left[
1+\ln\frac{\al(\nu_s)}{\al(\nu)}
\right]
\,.
\eea
In order to obtain Eq. (\ref{Vrexpanded}), we have used the explicit $\al(r^{-1})$ dependence of the potential. At the end we 
have a double expansion in $\al(\nu_s)$ and $\al(\nu)$. In the bound state calculation they get replaced by 
$\nu_s \sim m\al$ and $\nu \sim m\al^2$. 

We take Eq. (\ref{Vr1}) or Eq. (\ref{Vrexpanded}) as our final expressions 
for $V_{r,\MS}^{(2),RG}(r;\nu)$.

We have then obtained the RG improved and initial matching coefficients for the different potentials that 
compose $V_s$ to the order of interest. As we have already mentioned the $1/m$ and $1/m^2$ potentials 
suffer from field redefinitions ambiguities. Therefore, in some circumstances it can be convenient to cast the
initial matching conditions of $V_s$ in the following unified form\footnote{Note that we 
use $V_{r,\MS}^{(2)}(r)$ and not $V_{r,\MS}^{(2)}(r,\nu_s)$.}:
\be
V_{s,\MS}(r;\nu_s)=V^{(0)}_{s,\MS}(r;\nu_s)
+
\frac{V^{(1)}_{s,\MS}(r;\nu_s)}{m}+
\frac{1}{m^2}
\left(
{1 \over 2}\left\{{\bf p}^2,V_{{\bf p}^2,\MS}^{(2)}(r;\nu_s)\right\}
 + V_{r,\MS}^{(2)}(r)
\right)
\,,
\ee
as well as the total potential to be introduced in the Schroedinger equation
\bea
\label{VsMS}
V^{RG}_{s,\MS}(r;\nu)&=&V_{s,\MS}(r;\nu_s=1/r)+\delta V_{s,RG}(r;\nu_s=1/r,\nu)
\\
\nn
&=&
V^{(0),RG}_{s,\MS}(r;\nu)
+
\frac{V^{(1),RG}_{s,\MS}(r;\nu)}{m}+
\frac{1}{m^2}
\left(
{1 \over 2}\left\{{\bf p}^2,V_{{\bf p}^2,\MS}^{(2),RG}(r;\nu)\right\}
 + V_{r,\MS}^{(2),RG}(r;\nu)
\right)
\,.
\eea
Note that we have worked in a basis where $V_{L^2}$ is set to zero, at least within the accuracy of our computation. 

\section{Observables}
\label{Observables}
\subsection{Static potential and Energy}

The singlet static energy can be considered to be an observable for our purposes. 
It does not suffer from field redefinitions ambiguities and 
can be easily compared with the lattice determination.
It consists of the static potential, which is a Wilson coefficient, and an ultrasoft 
contribution, both of them can be taken either bare or renormalized. 
More explicitly, the expression for the static energy with NNNLL precision reads
\be
E_s(r)=V^{(0),RG}_{s,\MS}(r;\nu=\Delta V)+\delta E^{(0),us}_{s,\MS}(r;\nu=\Delta V)
\,,
\ee
where
\be
\delta E^{(0),us}_{s,\MS}(r,\nu)
=
-C_f {\bf r}^2(\Delta V)^3 V_A^2
\frac{\al(\nu) }{9 \pi } \left(6\ln \frac{\Delta V}{\nu}+6\ln 2-5\right)
\,.
\ee
Note that one has to be careful in using the same scheme in the soft and ultrasoft to obtain the correct result for the 
energy of the static singlet state. For ease of reference we also give the NNNLO result. It reads
\be
E_s^{NNNLO}(r)=V^{(0)}_{s,\MS}(r;\nu)+\delta E^{(0),us}_{s,\MS}(r;\nu)
\,.
\ee
We can also give the expression for the subleading ultrasoft contribution 
to the static energy:
\bea
&&
\delta E^{(0),us}_{s,\MS}|_{\rm NLO}
=
C_f {\bf r}^2(\Delta V)^3 V_A^2
\frac{\al^2(\nu) }{108 \pi ^2} 
\left(18 \beta_0 \ln ^2\left(\frac{\Delta V}{\nu
   }\right)
\right.
\\
\nn
&&
-6 \left(C_A \left(13+4 \pi ^2\right)-2 \beta_0 (-5+3\ln 2)\right) \ln \left(\frac{\Delta V}{\nu
   }\right)
   \\
   &&
   \nn
-2 C_A \left(-84+39 \ln 2+4 \pi ^2 (-2+3\ln 2)+72 \zeta (3)\right)+\beta_0 \left(67+3 \pi ^2-60 \ln 2+18\ln^2 2\right)\Bigg)
.
\eea 
This contribution is relevant for the complete 
N$^4$LO or N$^4$LL computation of the static energy. For the 
former the only computation left is the soft four-loop contribution to the static potential. 

\subsection{Energy $l\not=0$ $s=0$}

We are now in the position to obtain the heavy quarkonium energy with N$^3$LL accuracy (for $l\not=0$ and $s=0$):
\be
\label{Enjls}
E_{nljs}|_{l\not=0,s=0}=\left(E^{pot}_{\MS,nljs}+\delta E^{us}_{\MS,nl}\right)|_{l\not=0,s=0}
\,,
\ee
where $E^{pot}_{\MS,nljs}$ is the eigenvalue of the equation
\be
\label{EnpotMS}
\left(\frac{p^2}{2m_r}+V^{RG}_{s,\MS}\right)\phi_{njls}({\bf r})=E^{pot}_{\MS,nljs}\phi_{nljs}({\bf r})
\,,
\ee
and $V^{RG}_{s,\MS}$ is Eq. (\ref{VsMS}) minus $\left(\frac{1}{8m_1^3}+\frac{1}{8m_2^3}\right){\bf p}^4$.

The exact solution of Eq. (\ref{EnpotMS}) correctly produces all necessary soft and potential terms for the aimed  N$^3$LL accuracy, as well as some subleading terms. Such exact solution would only be possible to obtain through numerical methods (which on the other hand could actually be more easy to implement in practice). If we want to restrict ourselves to a strict N$^3$LL computation (in particular if seeking for an explicit analytical result), Eq. (\ref{EnpotMS}) should be computed within quantum mechanics perturbation theory up to NNNLO for general quantum numbers. Up to NNLO such computation was performed in Ref. \cite{Pineda:1997hz}. The lengthy N$^3$LO computation is missing, beyond the aim of this work, and will be considered elsewhere (for $l=0$ and $n=1$ such computation has been performed in Ref. \cite{Penin:2002zv}).

The ultrasoft correction to the energy due to the ultrasoft correction can be 
written in the following compact form
\bea
\label{deltaEusMSnl}
&&
\delta E^{us}_{\MS,nl}
=
\langle n,l |
\Biggl(
C_f {\bf r}(h_o-E_{n,l})^3 V_A^2
\left[
-\frac{\al }{9 \pi } \left(6 \ln \left(\frac{h_o-E_{n,l}}{\nu }\right)+6\ln 2-5\right)
\right.
\\
&& 
\nn  
+\frac{\al^2 }{108 \pi ^2} 
\left(18 \beta_0 \ln ^2\left(\frac{h_o-E_{n,l}}{\nu
   }\right)-6 \left(N_c \left(13+4 \pi ^2\right)-2 \beta_0 (-5+3\ln 2)\right) \ln \left(\frac{h_o-E_{n,l}}{\nu
   }\right)
   \right.
   \\
   &&
   \nn
   \left.\left.
+2C_A \left(84-39 \ln 2+4 \pi ^2 (2-3\ln 2)-72 \zeta (3)\right)+\beta_0 \left(67+3 \pi ^2-60 \ln 2+18\ln^2 2\right)\right)
   \right]{\bf r}
   \Biggr)
   | n,l \rangle 
\,,
\eea 
where the states $| n,l \rangle $ and the energies $E_{n,l}$ used above are the solution of the Schroedinger potential 
including the 1-loop static potential (i.e. with NLO accuracy): 
\be
\left[
\frac{{\bf p}^2}{2m_r}-\frac{C_f\,\alpha_s(\nu)}{r}\,
\bigg\{1+\frac{\alpha_s(\nu)}{4\pi} a_1(\nu;r)\bigg\}
\right]| n,l \rangle=E_{n,l}| n,l \rangle
\,.
\ee
$h_o$ could also be approximated to its NLO expression:
\be
h_o=\frac{{\bf p}^2}{2m_r}+\frac{1}{2N_c}\frac{\,\alpha_s(\nu)}{r}\,
\bigg\{1+\frac{\alpha_s(\nu)}{4\pi} a_1(\nu;r)\bigg\}
\,.
\ee
Eq. (\ref{deltaEusMSnl}) includes 
the complete LO ${\cal O}(m\al^5)$ and NLO  ${\cal O}(m\al^6)$ ultrasoft effects, as well as subleading effects. The LO expression would be 
enough for the N$^3$LL precision. A semianalytic expression exists for $l=0$ states \cite{Kniehl:1999ud} but missing for general quantum numbers 
and would require a dedicated study, again beyond the 
aim of this paper. 
In a strict fixed-order computation one should expand the wave functions 
to the appropriate order, as well as $h_o-E_{n,l}$, but  in some situations it could be more convenient to handle this expression numerically.

\section{Conclusions}
\label{conclusion}

We have computed the NLL  ultrasoft running 
of the spin-independent singlet potentials up to ${\cal O}(1/m^2)$, 
and the corresponding contribution to the spectrum. This includes 
the static energy at NNNLL order. 

Whereas there are still some pieces left for a complete NNNLL evaluation of the 
heavy quarkonium spectrum, our result provides with the missing link for the complete 
result for  $l\not=0$ and $s=0$ states, the structure of 
which is given in this paper for the first time. The full explicit analytic form will be presented elsewhere.

Another by-product of our computation is the  N$^4$LO ultrasoft spin-independent 
contribution to the heavy quarkonium mass and static energy.

\medskip

Note added: After our paper appeared on the web, the preprint \cite{Hoang:2011gy} was sent to the arXives. In this reference
expressions for the NLL ultrasoft running of the $1/m$ and $1/m^2$ potentials were given in the vNRQCD framework and agreement with
our results claimed. Note that in order the expressions for the $1/m^2$ potentials to be equal, the contribution associated to Eq. 31 in \cite{Hoang:2011gy} has to be added to the result obtained in Ref. \cite{Hoang:2006ht}. It also remains to be explained how (and if) the dependence on the infrared regulator mentioned in Ref. \cite{Max2} has disappeared.

\bigskip

\acknowledgments{
This work was partially supported by the spanish 
grants FPA2007-60275 and FPA2010-16963, and by the catalan grant SGR2009-00894.
}

\appendix
\section{Constants}
\be
T_F= {1 \over 2}; \quad C_A=N_c; \quad C_f = {N_c^2 - 1 \over 2\,N_c}\,.
\ee
\be
\beta_0=11\frac{C_A}{3}-{4 \over 3}T_Fn_f;
\qquad
\beta_1=34\frac{C_A^2}{3}-{20 \over 3}C_AT_Fn_f-4C_fT_Fn_f
;\ee
\be
\beta_2= {2857 \over 54}C_A^3-{1415 \over 27}C_A^2T_Fn_f+\frac{158}{27}C_AT_F^2n_f^2
-\frac{205}{9}C_AC_fT_Fn_f+\frac{44}{9}C_fT_F^2n_f^2+2C_f^2T_Fn_f
.\ee
\be
a_1={31C_A-20T_Fn_f \over 9};
\ee
\bea
\nonumber
&a_2 =& 
{{400\,{{{\it n_f}}^2}\,{{{\it T_F}}^2}}\over {81}} -
     {\it C_f}\,{\it n_f}\,{\it T_F}\,
      \left( {{55}\over 3} - 16\,\zeta(3) \right) 
\\
&&
\nn
 +
     {{{\it C_A}}^2}\,\left( {{4343}\over {162}} +
        {{16\,{{\pi }^2} - {{\pi }^4}}\over 4} + {{22\,\zeta(3)}\over 3}
        \right)
 - {\it C_A}\,{\it n_f}\,{\it T_F}\,
      \left( {{1798}\over {81}} + {{56\,\zeta(3)}\over 3} \right)
	  ;
\eea
\begin{eqnarray}
  a_3 = a_3^{(3)} n_f^3 + a_3^{(2)} n_f^2 + a_3^{(1)} n_f + a_3^{(0)}
  \,,
\end{eqnarray}
where
\begin{eqnarray}
  a_3^{(3)} &=& - \left(\frac{20}{9}\right)^3 T_F^3
  \,,\nonumber\\
  a_3^{(2)} &=&
  \left(\frac{12541}{243}
    + \frac{368\zeta(3)}{3}
    + \frac{64\pi^4}{135}
  \right) C_A T_F^2
  +
  \left(\frac{14002}{81}
    - \frac{416\zeta(3)}{3}
  \right) C_f T_F^2
  \,,\nonumber\\
  a_3^{(1)} &=&
  \left(-709.717
  \right) C_A^2 T_F
  +
  \left(-\frac{71281}{162}
    + 264 \zeta(3)
    + 80 \zeta(5)
  \right) C_AC_f T_F
  \nonumber\\&&\mbox{}
  +
  \left(\frac{286}{9}
    + \frac{296\zeta(3)}{3}
    - 160\zeta(5)
  \right) C_f^2 T_F
 +
  \left(-56.83(1)
  \right) \frac{d_F^{abcd}d_F^{abcd}}{N_A} \,,
  \nonumber\\
  a_3^{(0)} &=&
  502.24(1) \,\, C_A^3
  -136.39(12)\,\, \frac{d_F^{abcd}d_A^{abcd}}{N_A}
  \,,
  \label{eq::a3}
\end{eqnarray}
and
\be
\frac{d^{abcd}_F d^{abcd}_A}{N_A}=\frac{N_c(N_c^2+6)}{48}
\,.
\ee

The NRQCD matching coefficients have been computed over the years. In 
Ref. \cite{Manohar:1997qy} one can find the NRQCD matching coefficients of the 
one heavy quark sector at NLO, whereas in Ref. \cite{Pineda:1998kj} one can find them 
for the two heavy quark sector (although with a difference of scheme for $d_{vv}$). 
In Ref. \cite{BM} the LL soft running of the one heavy quark sector was obtained and 
in Ref. \cite{Pineda:2001ra} the LL soft running for the two heavy quark sector. $\al(m)$ has 
$n_f$ active light flavours:
\be
d_2(m)=\frac{\al(m)}{60\pi}T_F
,
\ee
\be
c_F(m)=1+\frac{\al(m)}{2\pi}(C_f+C_A)
,
\ee
\be
c_D(m)=1+\frac{\al(m)}{2\pi}C_A-16d_2(m)
.
\ee
\bea
d_{ss}^{a}(m)&=& \al^2(m) C_{f}\left({C_{A}\over 2}-C_{f}\right)
               \left(2-2\ln2 + i \pi\right) \,,\\
d_{sv}^{a}(m)&=& 0 \,,\\
d_{vs}^{a}(m)&=& {\al^2(m) \over 2}
               \left(-{3 \over 2}C_A+4C_f \right)
                \left(2-2\ln2 + i \pi\right) \,,\\
d_{vv}^{a}(m)&=& -\pi\al(m)\Biggl[1+{\al(m)\over\pi}\Biggl(
T_{R} \left[{1\over 3}n_{f}\left( 2\ln2-{5\over 3}-i\pi\right)
-{8\over
9}\right] \nonumber\\
&  &+C_{A}{109\over 36} - 4C_{f}\Biggr)\Biggr]
\,.
\eea

\bea
\nonumber
d_{ss}(m) &=& -{d_{ss}^a(m) \over 2N_c}- {3 d_{sv}^a(m)  \over 2N_c}
      -{N^2_c-1 \over 4N_c^2}d_{vs}^a(m)  - 3{N^2_c-1\over 4N_c^2} d_{vv}^a(m)
  + {2 \over 3}
C_{f}\left( {C_{A}\over 2}-C_{f}\right)
 \al^2(m)
\,,\\
\nonumber
d_{sv}(m)  &=& -{d_{ss}^a(m)  \over 2N_c}+ { d_{sv}^a(m)  \over 2N_c}
      -{N^2_c-1 \over 4N_c^2}d_{vs}^a(m)  + {N^2_c-1\over 4N_c^2} d_{vv}^a(m) +
C_{f}\left( {C_{A}\over 2}-C_{f}\right)
 \al^2(m)
\,,\\
\nonumber
d_{vs}(m)  &=& -d_{ss}^a(m)  - 3 d_{sv}^a(m) 
      +{d_{vs}^a(m)  \over 2N_c}+  {3 d_{vv}^a(m)  \over 2N_c}
+ \left({4 \over 3} C_f 
  + {11 \over 12} C_A \right)\al^2(m)
\,,\\
d_{vv}(m)  &=&  -d_{ss}^a(m)  +  d_{sv}^c(m) 
      +{d_{vs}^a(m)  \over 2N_c}- { d_{vv}^a(m)  \over 2N_c}
+\al^2(m)(2C_f-\frac{C_A}{2})
\,.\eea

We now 
define $z=\left[{\al(\nu_s) \over \al(m)}\right]^{1 \over
\beta_0}\simeq 1 -1/(2\pi)\al(\nu_s)\ln ({\nu_s \over m})$,
\bea
c_F(\nu_s)&=&c_F(m)-1+z^{-C_A}
\,,
\nn\\
c_D(\nu_s)&=&c_D(m)-1+
{9C_A \over 9C_A+8T_Fn_f}
\left\{
-\frac{5 C_A + 4 T_F n_f}{4 C_A + 4
T_F n_f} z^{-2 C_A} +
\frac{C_A +16 C_f - 8 T_F n_f}{2(C_A-2T_F n_f)}
\right.
\nn
\\
&&\qquad
+ \frac{-7 C_A^2 + 32 C_A C_f - 4 C_A T_F n_f +32 C_f T_F
n_f}{4(C_A + T_F n_f)(2 T_F n_f-C_A)} z^{4 T_F n_f/3 - 2C_A/3}
\nn
\\
&&
\qquad
\left.
+{8T_Fn_f \over 9C_A}
\left[
z^{-2C_A}+\left({20 \over 13}+{32 \over 13}{C_f \over
C_A}\right)\left[1-z^{-13C_A \over 6}\right]
\right]
\right\}
\,,
\nn\\
d_{ss}(\nu_s)&=&d_{ss}(m)+4C_f\left(C_f-{C_A \over 2}\right){\pi\over\beta_0}\al(m)\left[z^{\beta_0}-1\right]
\,,
\nn\\
d_{sv}(\nu_s)&=&d_{sv}(m)
\,,
\nn\\
d_{vs}(\nu_s)&=&d_{vs}(m)-\left(C_f-C_A\right){8\pi\over\beta_0}\al(m)
\left[z^{\beta_0}-1 \right]
\nn
\\
&&
-{27C_A^2 \over 9C_A+8T_Fn_f}{\pi \over \beta_0}\al(m)
\left\{
-\frac{5 C_A + 4 T_F n_f}{4 C_A + 4
T_F n_f}{\beta_0 \over \beta_0-2C_A}\left(z^{\beta_0-2 C_A}-1\right) 
\right.
\nn
\\
&&
\qquad
+
\frac{C_A +16 C_f - 8 T_F n_f}{2(C_A-2T_F n_f)}\left(z^{\beta_0}-1\right)
\nn
\\
&&
\qquad
+ \frac{-7 C_A^2 + 32 C_A C_f - 4 C_A T_F n_f +32 C_f T_F
n_f}{4(C_A + T_F n_f)(2 T_F n_f-C_A)}
\nn
\\
&&
\qquad
\qquad
\times
{3\beta_0 \over 3\beta_0+4T_Fn_f-2C_A}\left( z^{\beta_0+4 T_F n_f/3 - 2C_A/3}-1\right)
\nn
\\
&&
\qquad
+{8T_Fn_f \over 9C_A}
\left[{\beta_0 \over \beta_0-2C_A}\left(z^{\beta_0-2C_A}-1\right)
+\left({20 \over 13}+{32 \over 13}{C_f \over C_A}\right)
\right.
\nn
\\
&&
\qquad\qquad
\left.
\left.
\times
\left(
\left[z^{\beta_0}-1\right]-{6\beta_0 \over 6\beta_0-13C_A}
\left[z^{\beta_0-{13C_A \over 6}}-1\right]
\right)
\right]
\right\}
\,,
\nn\\
d_{vv}(\nu_s)&=&d_{vv}(m)+{C_A \over
\beta_0-2C_A}\pi\al(m)\left\{z^{\beta_0-2C_A}-1\right\}
\label{RGeqhs}
\,.
\eea
The results displayed above for the NRQCD matching coefficients, $c$'s and $d$'s,
are correct with LL and NLO accuracy, but not beyond. 

We will also need
\bea
c_1^{hl}(\nu_s)&=&
{9C_A \over 9C_A+8T_Fn_f}
\left\{
\frac{5 C_A + 4 T_F n_f}{4 C_A + 4
T_F n_f} z^{-2 C_A} -
\frac{C_A +16 C_f - 8 T_F n_f}{2(C_A-2T_F n_f)}
\right.
\nn
\\
&&\qquad
- \frac{-7 C_A^2 + 32 C_A C_f - 4 C_A T_F n_f +32 C_f T_F
n_f}{4(C_A + T_F n_f)(2 T_F n_f-C_A)} z^{4 T_F n_f/3 - 2C_A/3}
\nn
\\
&&
\qquad
\left.
+
z^{-2C_A}+\left({20 \over 13}+{32 \over 13}{C_f \over
C_A}\right)\left[1-z^{-13C_A \over 6}\right]
\right\}
\,.
\eea
This is correct with LL accuracy, which is enough for us.


\begin{references}

\bibitem{Anzai:2009tm}
  C.~Anzai, Y.~Kiyo, Y.~Sumino,
  Phys.\ Rev.\ Lett.\  {\bf 104}, 112003 (2010).
  [arXiv:0911.4335 [hep-ph]].

\bibitem{Smirnov:2009fh}
  A.~V.~Smirnov, V.~A.~Smirnov, M.~Steinhauser,
  Phys.\ Rev.\ Lett.\  {\bf 104}, 112002 (2010).
  [arXiv:0911.4742 [hep-ph]].
 
\bibitem{Kniehl:2002br}
  B.~A.~Kniehl, A.~A.~Penin, V.~A.~Smirnov and M.~Steinhauser,
  Nucl.\ Phys.\  B {\bf 635}, 357 (2002)
  [arXiv:hep-ph/0203166].

\bibitem{Penin:2002zv}
  A.~A.~Penin and M.~Steinhauser,
  Phys.\ Lett.\  B {\bf 538}, 335 (2002)
  [arXiv:hep-ph/0204290].

\bibitem{Kniehl:2003ap}
  B.~A.~Kniehl, A.~A.~Penin, A.~Pineda, V.~A.~Smirnov and M.~Steinhauser,
  Phys.\ Rev.\ Lett.\  {\bf 92}, 242001 (2004)
  [Erratum-ibid.\  {\bf 104}, 199901 (2010)]
  [arXiv:hep-ph/0312086].

\bibitem{Penin:2004xi}
  A.~A.~Penin, A.~Pineda, V.~A.~Smirnov and M.~Steinhauser,
  Phys.\ Lett.\  B {\bf 593}, 124 (2004)
  [Erratum-ibid.\  {\bf 677}, 343 (2009)]
  [Erratum-ibid.\  {\bf 683}, 358 (2010)]
  [arXiv:hep-ph/0403080].

\bibitem{Brambilla:2009bi}
  N.~Brambilla, A.~Vairo, X.~Garcia i Tormo and J.~Soto,
  Phys.\ Rev.\  D {\bf 80}, 034016 (2009)
  [arXiv:0906.1390 [hep-ph]].

\bibitem{Pineda:2010mb}
  A.~Pineda and M.~Stahlhofen,
  Phys.\ Rev.\  D {\bf 81}, 074026 (2010)
  [arXiv:1002.1965 [hep-th]].

\bibitem{Pineda:2007kz}
  A.~Pineda,
  Phys.\ Rev.\  D {\bf 77}, 021701 (2008)
  [arXiv:0709.2876 [hep-th]].

\bibitem{Koma:2006si}
  Y.~Koma, M.~Koma and H.~Wittig,
  Phys.\ Rev.\ Lett.\  {\bf 97}, 122003 (2006).
  
\bibitem{Koma:2006fw}
  Y.~Koma and M.~Koma,
  Nucl.\ Phys.\  B {\bf 769}, 79 (2007).
  
\bibitem{Brambilla:2000gk}
  N.~Brambilla, A.~Pineda, J.~Soto and A.~Vairo,
  Phys.\ Rev.\  D {\bf 63}, 014023 (2000)
  [arXiv:hep-ph/0002250].

\bibitem{Pineda:2000sz}
  A.~Pineda and A.~Vairo,
  Phys.\ Rev.\  D {\bf 63}, 054007 (2001)
  [Erratum-ibid.\  D {\bf 64}, 039902 (2001)]
  [arXiv:hep-ph/0009145].

\bibitem{Hoang:2006ht}
  A.~H.~Hoang and M.~Stahlhofen,
  Phys.\ Rev.\  D {\bf 75}, 054025 (2007)
  [arXiv:hep-ph/0611292].

\bibitem{Luke:1999kz}
  M.~E.~Luke, A.~V.~Manohar and I.~Z.~Rothstein,
  Phys.\ Rev.\  D {\bf 61}, 074025 (2000)
  [arXiv:hep-ph/9910209].

\bibitem{Manohar:1999xd}
  A.~V.~Manohar and I.~W.~Stewart,
  Phys.\ Rev.\  D {\bf 62} (2000) 014033
  [arXiv:hep-ph/9912226].

\bibitem{Hoang:2002yy}
  A.~H.~Hoang and I.~W.~Stewart,
  Phys.\ Rev.\  D {\bf 67}, 114020 (2003)
  [arXiv:hep-ph/0209340].

\bibitem{Max2}
  M.~Stahlhofen, PhD thesis, 
  ``Ultrasoft renormalization of the potentials in vNRQCD''.

\bibitem{Pineda:1997bj}
  A.~Pineda and J.~Soto,
  Nucl.\ Phys.\ Proc.\ Suppl.\  {\bf 64}, 428 (1998).

\bibitem{Brambilla:1999xf}
  N.~Brambilla, A.~Pineda, J.~Soto and A.~Vairo,
  Nucl.\ Phys.\  B {\bf 566}, 275 (2000)
  [arXiv:hep-ph/9907240].

\bibitem{Pineda:2006ri}
  A.~Pineda and A.~Signer,
  Nucl.\ Phys.\  B {\bf 762}, 67 (2007)
  [arXiv:hep-ph/0607239].

\bibitem{Pineda:2000gz}
  A.~Pineda and J.~Soto,
  Phys.\ Lett.\  B {\bf 495}, 323 (2000).
    
\bibitem{Pineda:1997ie}
  A.~Pineda and J.~Soto,
  Phys.\ Lett.\  B {\bf 420}, 391 (1998).

\bibitem{Brambilla:1999qa}
  N.~Brambilla, A.~Pineda, J.~Soto and A.~Vairo,
  Phys.\ Rev.\  D {\bf 60}, 091502 (1999).
 
\bibitem{Kniehl:1999ud}
  B.~A.~Kniehl and A.~A.~Penin,
  Nucl.\ Phys.\  B {\bf 563}, 200 (1999).

\bibitem{Eidemuller:1997bb}
  M.~Eidemuller and M.~Jamin,
  Phys.\ Lett.\  B {\bf 416}, 415 (1998)
  [arXiv:hep-ph/9709419].

\bibitem{Brambilla:2006wp}
  N.~Brambilla, X.~Garcia i Tormo, J.~Soto and A.~Vairo,
  Phys.\ Lett.\  B {\bf 647}, 185 (2007).

\bibitem{MaxHyb}
  A.~Pineda and M.~Stahlhofen,
  arXiv:1105.4356 [hep-ph].

\bibitem{Schroder:1999sg}
  Y.~Schroder,
  ``The static potential in QCD'', DESY-THESIS-1999-021.

  \bibitem{FSP} W. Fischler, Nucl. Phys. {\bf B129}, 157 (1977);
Y. Schr\"oder, Phys. Lett. {\bf B447}, 321 (1999).

\bibitem{Kniehl:2001ju}
  B.~A.~Kniehl, A.~A.~Penin, M.~Steinhauser and V.~A.~Smirnov,
  Phys.\ Rev.\  D {\bf 65} (2002) 091503
  [arXiv:hep-ph/0106135].

\bibitem{Pineda:1998kj}
  A.~Pineda and J.~Soto,
  Phys.\ Rev.\  D {\bf 58}, 114011 (1998)
  [arXiv:hep-ph/9802365].

\bibitem{Manohar:2000hj}
  A.~V.~Manohar and I.~W.~Stewart,
  Phys.\ Rev.\  D {\bf 62}, 074015 (2000)
  [arXiv:hep-ph/0003032].

\bibitem{Pineda:1997hz}
  A.~Pineda and F.~J.~Yndurain,
  Phys.\ Rev.\  D {\bf 58}, 094022 (1998)
  [arXiv:hep-ph/9711287].

\bibitem{Hoang:2011gy}
  A.~H.~Hoang and M.~Stahlhofen,
  arXiv:1102.0269 [hep-ph].

\bibitem{Manohar:1997qy}
  A.~V.~Manohar,
  Phys.\ Rev.\  D {\bf 56}, 230 (1997)
  [arXiv:hep-ph/9701294].

\bibitem{BM}  C. Bauer and A.V. Manohar, Phys. Rev. {\bf D57}, 337 (1998).

\bibitem{Pineda:2001ra}
  A.~Pineda,
  Phys.\ Rev.\  D {\bf 65}, 074007 (2002)
  [arXiv:hep-ph/0109117].
    
\end{references}
\end{document}